\documentclass[prl,onecolumn,lengthcheck]{revtex4-2}

\usepackage{amssymb, amsmath, amsthm, amstext, amsfonts, amsopn, amscd}
\usepackage{mathrsfs, dsfont, mathdots}
\usepackage{bm}
\usepackage{graphicx}
\usepackage{xcolor}
\usepackage[colorlinks=true, linkcolor=black, breaklinks=true, citecolor=blue, urlcolor=red]{hyperref} 
\usepackage{float}
\usepackage{subcaption}
\usepackage{caption}
\usepackage{tikz}
\usepackage{dcolumn}
\usepackage{todonotes}

\theoremstyle{definition}

\theoremstyle{remark}

\newcommand{\tr}{\operatorname{tr}}
\newcommand{\ptr}{\operatorname{ptr}}

\newcommand{\fA}{\mathfrak A}

\newcommand{\C}{\mathbb{C}}

\newcommand{\cE}{\mathcal E}
\newcommand{\fF}{\mathfrak{F}}

\newcommand{\fh}{\mathfrak h}
\newcommand{\cH}{\mathcal H}
\newcommand{\ltwo}{\ell^{2}}

\newcommand{\cP}{\mathcal{P}}
\newcommand{\R}{\mathbb{R}}
\newcommand{\cS}{\mathcal S}

\newcommand{\Z}{\mathbb{Z}}

\newcommand{\1}{\mathds{1}}

\newcommand{\vep}{\varepsilon}

\DeclareMathOperator{\sign}{sign}

\DeclareMathOperator{\Ad}{Ad}

\DeclareMathOperator{\CAR}{CAR}

\begin{document}

\title{On the renormalization group fixed point of the two-dimensional Ising model at criticality}

\author{Alexander Stottmeister$^{1}$ and Tobias J.\ Osborne$^{1}$}

\affiliation{$^{1}$Institut f\"ur Theoretische Physik, Leibniz Universit\"at Hannover, Appelstr. 2, 30167 Hannover, Germany}

\begin{abstract}
We analyze the renormalization group fixed point of the two-dimensional Ising model at criticality. In contrast with expectations from tensor network renormalization (TNR), we show that a simple, explicit analytic description of this fixed point using operator-algebraic renormalization (OAR) is possible. Specifically, the fixed point is characterized in terms of spin-spin correlation functions. Explicit error bounds for the approximation of continuum correlation functions are given.
\end{abstract}

\maketitle

\paragraph{Introduction.---}\hspace{-1em}The statistical mechanics of classical lattice systems continue to present fascinating and remarkable physics. The stochastic geometry exhibited by models as fundamental and elementary as the Ising model \cite{isingBeitragZurTheorie1925} exhibits a beautiful structure whose active study persists to the current day \cite{grimmettRandomClusterModel2006}. Most intriguing here is the critical phenomena of the model as it approaches a phase transition \cite{schroederIntroductionThermalPhysics2000}. Applications of the Ising model and its generalisations range from superconductivity \cite{liRecentProgressesTwodimensional2021}, fault-tolerant quantum computation \cite{pachosIntroductionTopologicalQuantum2012a}, high energy physics \cite{iqbal3dIsingModel2020}, to genetics \cite{majewskiIsingModelPhysics2001,krishnanModifiedIsingModel2020} and the social sciences \cite{lipowskiPhaseTransitionPowerlaw2017} and beyond. The two-dimensional case of the Ising model is one of the most well-studied systems in statistical physics, with nearly 80 years of history dating back at least to 1944, with the celebrated work of Lars Onsager \cite{onsagerCrystalStatisticsTwoDimensional1944}, who solved the the model on a square lattice in the absence of external magnetic field. This solution is the cornerstone of much of modern statistical physics, and thereby the Ising model has become the benchmark for analytic and numerical methods alike. 

During the past decade tensor networks \cite{bridgemanHandwavingInterpretiveDance2017} have risen to prominence as a powerful tool to study complex systems. These have a rich history originating in the works of Kadanoff \cite{kadanoffNotesMigdalRecursion1976,KadanoffScalingLawsFor} and Wilson \cite{WilsonTheRenormalizationGroupKondo,WilsonTheRenormalizationGroupCritical}, the density matrix renormalization group \cite{WhiteDensityMatrixFormulation}, and branching out into a multitude of methods with a wide variety of applications from 2D systems through to models with anyonic excitations.
One fascinating area of such works applies modern tensor-network techniques to classical models of statistical physics. This was arguably revolutionized by the tensor renormalization group (TRG) of Levin-Nave \cite{LevinTensorRenormalizationGroup} having a wide range of applications \cite{zhaoRenormalizationTensornetworkStates2010, dittrichCoarseGrainingMethods2012, kadohInvestigationComplexPh42020, kuramashiTensorRenormalizationGroup2020}, which has been refined in various forms, in particular to deal with entanglement of local degrees of freedom such as tensor entanglement-filtering renormalization group \cite{guTensorentanglementfilteringRenormalizationApproach2009a,guTensorentanglementRenormalizationGroup2008}, high-order tensor renormalization group \cite{xieSecondRenormalizationTensorNetwork2009, xieCoarsegrainingRenormalizationHigherorder2012, bazavovTensorRenormalizationGroup2019, buttTensorNetworkFormulation2020, blochTensorRenormalizationGroup2021, liTensornetworkRenormalizationApproach2022}, tensor network renormalization (with or without positivity) \cite{EvenblyTensorNetworkRenormalization, balRenormalizationGroupFlows2017, yangLoopOptimizationTensor2017}.
Here impressive numerical results suggest the general applicability of the TRG, and relatives such as tensor network renormalization, as a general purpose method for investigating partition functions of classical lattice models. Although the TRG does flow to a fixed point off criticality -- i.e., an infinite bond dimension is required to express the fixed-point tensor -- it is still useful for the study of critical phenomena. The goal of explicitly computing fixed-point tensors for critical systems -- closely related to the approximation of continuum limits -- is still an outstanding challenge for tensor-network methods.

The desire for an explicit RG capable of describing the continuum limit of lattice discretizations of quantum field theories has led to the recent development of operator algebraic renormalization (OAR) \cite{BrothierAnOperatorAlgebraic,BrothierConstructionsOfConformal,MorinelliScalingLimitsOf,OsborneCFTapprox,OsborneCFTsim,StottmeisterOperatorAlgebraicRenormalization,OsborneContinuumLimitsOf, StottmeisterAnyonBraidingAnd}. This emerging RG method is closely related to tensor network methods such as the multi-scale entanglement renormalization ansatz (MERA) \cite{vidalEntanglementRenormalization2007,vidalEntanglementRenormalizationIntroduction2011,EvenblyTNRMERA}, and has enjoyed notable recent successes in the computational and analytic approximation of a variety of quantum field theories, from conformal field theories to higher-dimensional models. It is an intriguing open question to determine whether OAR is applicable in the context of classical criticality and, if so, whether it can furnish any information about the fixed-point tensor at phase transitions. 

In this Letter we demonstrate that OAR is capable of exactly representing critical points of classical lattice models. To do this we generalize OAR to apply to partition functions of classical lattice models and analytically compute the action of the OAR group on the transfer operator of the 2D Ising model. We obtain thereby an explicit and analytic representation of the fixed-point tensor. In accordance with expectations arising in previous TRG studies we find that this tensor requires an infinite bond dimension.

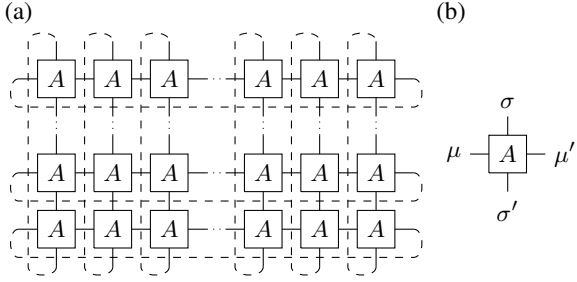
\begin{figure}[ht]
\centering
	\begin{tikzpicture}
		\draw (-0.25,3.125) node{(a)};
		
		\foreach \i in {0,0.75,1.5,2.75,3.5,4.25}
		{
		\foreach \j in {-0.75,0,0.75,2}
		{
		\draw (0.25+\i,0.5+\j) to (0.25+\i,0.75+\j);
		}
		\foreach \j in {0,0.75,2}
		{
		\draw (0.5+\i,0.25+\j) to (0.75+\i,0.25+\j);
		}
		\foreach \j in {0,0.75,2}
		{
		\draw (0+\i,0+\j) to (0.5+\i,0+\j) to (0.5+\i,0.5+\j) to (0+\i,0.5+\j) to (0+\i,0+\j);
		\draw (0.25+\i,0.25+\j) node{$A$};
		}
		}
		\foreach \j in {0,0.75,2}
		{
		\draw (-0.25,0.25+\j) to (0,0.25+\j);
		\draw[dotted] (2.25,0.25+\j) to (2.5,0.25+\j);
		\draw (2.5,0.25+\j) to (2.75,0.25+\j);
		}
		\foreach \j in {0,0.75,2}
		{
		\draw[dashed] (-0.25,0.25+\j) to[out=180,in=180] (-0.25,-0.125+\j) to (5,-0.125+\j) to[out=0,in=0] (5,0.25+\j);
		}
		\foreach \i in {0,0.75,1.5,2.75,3.5,4.25}
		{
		\draw[dashed] (0.25+\i,-0.25) to[out=270,in=270] (-0.125+\i,-0.25) to (-0.125+\i,2.75) to[out=90,in=90] (0.25+\i,2.75);
		\draw[dotted] (0.25+\i,1.5) to (0.25+\i,1.75);
		\draw (0.25+\i,1.75) to (0.25+\i,2);
		}
		
		\draw (5.5,3.125) node{(b)};
		
		\draw (6,1) to (6.5,1) to (6.5,1.5) to (6,1.5) to (6,1) (6.25,1.25) node{$A$};
		\draw (5.75,1.25) to (6,1.25) (5.75,1.25) node[left]{$\mu$};
		\draw (6.5,1.25) to (6.75,1.25) (6.75,1.25) node[right]{$\mu'$};
		\draw (6.25,1.5) to (6.25,1.75) (6.25,1.75) node[above]{$\sigma$};
		\draw (6.25,1) to (6.25,0.75) (6.25,0.75) node[below]{$\sigma'$};
	\end{tikzpicture}
	\caption{Illustration of the partition function $Z_{MN}$ in (a) as a two-dimensional tensor network built from the local tensor $A$ in (b). Dashed lines indicate contractions due to periodic boundary conditions.}
	\label{fig:tn}
\end{figure}

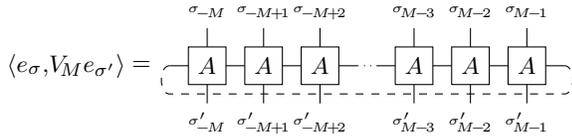
\begin{figure}[ht]
\centering
	\begin{tikzpicture}
		\draw (-1.625,0.25) node{$\langle e_{\sigma},\!V_{\!M}e_{\sigma'}\rangle$} (-0.625,0.25) node{$=$};
		\foreach \i in {0,0.75,1.5,2.75,3.5,4.25}
		{
		\foreach \j in {-0.75,0}
		{
		\draw (0.25+\i,0.5+\j) to (0.25+\i,0.75+\j);
		}
		\draw (0.5+\i,0.25) to (0.75+\i,0.25);
		\draw (0+\i,0) to (0.5+\i,0) to (0.5+\i,0.5) to (0+\i,0.5) to (0+\i,0);
		\draw (0.25+\i,0.25) node{$A$};
		}
		\draw (-0.25,0.25) to (0,0.25);
		\draw[dotted] (2.25,0.25) to (2.5,0.25);
		\draw (2.5,0.25) to (2.75,0.25);
		\draw[dashed] (-0.25,0.25) to[out=180,in=180] (-0.25,-0.125) to (5,-0.125) to[out=0,in=0] (5,0.25);
		\draw (0.25,0.75) node[above]{\tiny$\sigma_{\!-\!M}$};
		\draw (1,0.75) node[above]{\tiny$\sigma_{\!-\!M\!+\!1}$};
		\draw (1.75,0.75) node[above]{\tiny$\sigma_{\!-\!M\!+\!2}$};
		\draw (3,0.75) node[above]{\tiny$\sigma_{\!M\!-\!3}$};
		\draw (3.75,0.75) node[above]{\tiny$\sigma_{\!M\!-\!2}$};
		\draw (4.5,0.75) node[above]{\tiny$\sigma_{\!M\!-\!1}$};
		
		\draw (0.25,-0.25) node[below]{\tiny$\sigma'_{\!-\!M}$};
		\draw (1,-0.25) node[below]{\tiny$\sigma'_{\!-\!M\!+\!1}$};
		\draw (1.75,-0.25) node[below]{\tiny$\sigma'_{\!-\!M\!+\!2}$};
		\draw (3,-0.25) node[below]{\tiny$\sigma'_{\!M\!-\!3}$};
		\draw (3.75,-0.25) node[below]{\tiny$\sigma'_{\!M\!-\!2}$};
		\draw (4.5,-0.25) node[below]{\tiny$\sigma'_{\!M\!-\!1}$};
	\end{tikzpicture}
	\caption{Illustration of the Horizontal transfer matrix $V_{\!M}$ associated with the tensor $A$.}
	\label{fig:tm}
\end{figure}

\paragraph{Basics of 2d Ising.}
The two-dimensional anisotropic Ising model on a $N\times M$ square lattice with periodic boundary conditions can be naturally formulated as a tensor network (see Fig.~\ref{fig:tn}), i.e.~its canonical partition,
\begin{align}
\label{eq:2ditn}
Z_{MN} &\!=\!\sum_{\{\sigma\},\{\mu\}}\prod_{k=-N}^{N-1}\prod_{j=-M}^{M-1}A_{\mu_{j,k}\mu_{j+1,k}\sigma_{j,k}\sigma_{j,k+1}}, 
\end{align}
is given in terms of the tensor,
\begin{align}
\label{eq:2dit}
A_{\mu\mu'\sigma\sigma'} &\!=\!\delta_{\mu.\sigma}e^{K_{1}\mu\mu'}e^{K_{2}\sigma\sigma'},
\end{align} 
with $\mu,\mu',\sigma,\sigma'\in\{\pm1\}$ as well as horizontal and vertical coupling constants $K_{1},K_{2}$.
Spin-spin and other correlation functions are conveniently expressed using the horizontal transfer matrix $V_{M}$ (see Fig.~\ref{fig:tm}) naturally given in the $\sigma^{(3)}$-basis \cite{SchultzTwoDimensionalIsing, SatoHolonomicQuantumFieldsV}:
\begin{align}
\label{eq:2ditm}
\langle e_{\sigma},\!\!V_{\!M}e_{\sigma'}\!\rangle &\!=\!\!\sum_{\{\mu\}}\!\prod_{j=-\!M}^{M-1}\!\!\!\!A_{\mu_{j}\mu_{j\!+\!1}\sigma_{\!j}\sigma'_{\!j}}\!\!=\!\!\!\!\prod_{j=-\!M}^{M-1}\!\!\!\!e^{K_{\!1}\sigma_{\!j}\sigma_{\!j\!+\!1}÷\!+\!K_{\!2}\sigma_{\!j}\sigma'_{\!j}},
\end{align}
where $e_{\sigma}\!=\!\otimes_{j=-\!M}^{M}e_{\sigma_{j}}$, $\sigma^{(3)}_{j}e_{\sigma_{j}}\!=\!\sigma_{j}e_{\sigma_{j}}$. As an operator on the Hilbert space $\cH_{\!M}\!=\!\otimes_{j=-\!M}^{M}\C^{2}$, associated with each row of the lattice, the transfer matrix $V_{\!M}$ takes the form,
\begin{align}
\label{eq:2ditmop}
V_{\!M} &\!=\!C(2K_{2})^{\!\frac{M}{2}}\!e^{K_{1}\!\sum_{j=\!-\!M}^{M-2}\!\sigma^{\!(3)}_{j}\!\sigma^{\!(3)}_{j\!+\!1}}e^{K_{2}^{*}\!\sum_{j=\!-\!M}^{M-1}\!\sigma^{\!(1)}_{j}},
\end{align}
where $\tanh(K_{2}^{*})\!=\!e^{-2K_{2}}$ and $C(K_{2})\!=\!2\sinh(2K_{2})$, which decomposes into operators associated with vertical couplings, $V^{(1)}_{\!M}\!=\!(2\sinh(2K_{2})\!)^{\!\frac{M}{2}}e^{K_{2}^{*}\!\sum_{j=\!-\!M}^{M-1}\!\sigma^{\!(1)}_{j}}$, and horizontal coupling respectively $V^{(3)}_{\!M}\!=\!e^{K_{1}\!\sum_{j=\!-\!M}^{M-1}\!\sigma^{\!(3)}_{j}\!\sigma^{\!(3)}_{j\!+\!1}}$. While the partition function is given by the trace of the horizontal transfer matrix, $Z_{MN}\!=\!\tr(V_{\!M})$, the correlation functions are more naturally expressed using the symmetrized transfer matrix,
\begin{align}
\label{eq:2ditmsym}
V^{(\textup{sym})}_{\!M} & \!=\!\big(V^{(3)}_{\!M}\big)^{\frac{1}{2}}V^{(1)}_{\!M}\big(V^{(3)}_{\!M}\big)^{\frac{1}{2}},
\end{align}
which results in:
\begin{align}
\label{eq:2dicor}
\langle\sigma_{j_{1},k_{1}}...\sigma_{j_{n},k_{n}}\rangle & \!=\!\tfrac{1}{Z_{MN}}\tr\!\Big(\!\!\big(V^{(\textup{sym})}_{\!M}\big)^{\!N}\sigma^{(3)}_{j_{1}k_{1}}...\sigma^{(3)}_{j_{n}k_{n}}\!\Big),
\end{align}
where $\sigma^{(3)}_{jk}\!=\!\big(V^{(\textup{sym})}_{\!M}\big)^{\!k}\sigma^{(3)}_{j}\big(V^{(\textup{sym})}_{\!M}\big)^{\!-k}$.

\paragraph{OAR for 2d Ising.}
Exploiting the operator-algebraic structure of the transfer matrix formulation, we can apply OAR to analyze the large-scale behavior of correlation functions: $V^{(\textup{sym})}_{\!M}$ is a positive, trace-class operator on $\cH_{\!M}$ inducing a quasi-free Gibbs state, $\rho_{\!M\!N}\!=\!\tfrac{1}{Z_{\!M\!N}}\big(V^{(\textup{sym})}_{\!M}\big)^{\!N}$, on the quantum spin chains given in terms of the Pauli algebra $\cP_{\!M}\!=\!\otimes_{j=-\!M}^{M-1}M_{2}(\C)$. By the Jordan-Wigner transform \cite{EvansQuantumSymmetriesOn}, $a_{j}\!=\!\big(\prod_{-\!M\leq l<j}\sigma^{(1)}_{l}\big)\tfrac{1}{2}(\sigma^{(3)}_{j}\!\!+\!i\sigma^{(2)}_{j})$, the latter is isomorphic to the algebra of complex fermions $\fA_{\!M}\!=\!\fA_{\CAR}(\fh_{\!M})$ with one-particle Hilbert space $\fh_{\!M}\!=\!\ltwo(\Lambda_{\!M})$, $\Lambda_{M}\!=\!\{-M,...,M\!-\!1\}$\footnote{The boundary conditions for $\fh_{M}$ are chosen such that $V^{(3)}_{M}$ is an exponential of quadratic expressions in annihilation and creation operators \cite{SchultzTwoDimensionalIsing}. For finite $M$, the eigenstate corresponding to the largest eigenvalue of $V^{(\textup{sym})}_{\!M}$ is obtained with anti-periodic boundary conditions for $\fh_{M}$.}. We define the renormalization group transformation\footnote{A trace-preserving quantum channel.}, $\cE\!:\!\cS_{2M}\!\rightarrow\!\cS_{M}$, that coarse grains states on the chain of twice the length, $\cS_{2M}$, to those on the given length, $\cS_{M}$, by its dual quantum channel, $\alpha:\!\fA_{M}\rightarrow\!\fA_{2M}$:
\begin{align}
\label{eq:dual}
\tr(\cE(\rho)A) & \!=\! \tr(\rho\!\ \alpha(A)\!), & \rho\in\cS_{2M},\!\ A\in\fA_{M}.
\end{align}
The dual quantum channel is naturally given by an isometry \cite{OsborneCFTapprox}, $R:\fh_{M}\!\rightarrow\!\fh_{2M}$:
\begin{align}
\label{eq:rg1p}
\alpha(a(\xi)\!) & \!=\! a(R(\xi)\!),\! & \!R(\xi)_{j'} & \!=\!\!\sum_{j=-M}^{M-1}\!\!\xi_{j}\sum_{n\in\Z}h_{n}\delta_{2j,j'-n}
\end{align}
for $\xi\in\fh_{\!M}$ and $a(\xi)\!=\!\sum_{j=-M}^{M-1}\bar{\xi}_{j}a_{j}$. The coefficients $h_{n}$ are given by the low-pass filter of a real, orthonormal, compactly supported scaling function $s\!\in\!C^{r}(\R)$, satisfying the scaling equation $s(x)\!=\!\sum_{n\in Z}h_{n}2^{\frac{1}{2}}s(2x\!-\!n)$ (appropriately periodized to comply with the boundary conditions) \cite{DaubechiesTenLecturesOn}. The renormalization group transformation takes a particularly simple form in momentum space,
\begin{align}
\label{eq:rg1pft}
R(\hat{\xi})_{\theta'} & \!=\!2^{\frac{1}{2}}m_{0}(\theta')\hat{\xi}_{2\theta'}, &\!\!\! m_{0}(\theta') & \!=\!\tfrac{1}{\sqrt{2}}\!\sum_{n\in\Z}h_{n}e^{-i\theta'n},
\end{align}
where $\hat{\xi}_{\theta}\!=\!\sum_{j=-M}^{M-1}e^{-i\theta j}\xi_{j}$ for $\theta\!\in\!\tfrac{\pi}{M}\{-M,...,M\!-\!1\}$. In this way, we realize (discrete) renormalization group flow lines within the state space $\cS_{M}$ by,
\begin{align}
\label{eq:rgflow}
\rho^{(m)}_{MN} & \!=\! \cE^{m}(\rho_{(M+m)N}), & 0 & \!\leq\!m\!<\!\infty,
\end{align}
using the Gibbs state $\rho_{\!M\!N}$ as an input. On the Pauli algebra $\cP_{\!M}$, the coarse graining takes the form: $\cE(\!\ \cdot\!\ )\!=\!\ptr(U_{\!M}^{*}(\!\ \cdot\!\ )U_{\!M})$, where $\ptr$ is the partial trace with respect to the natural embedding $\cH_{\!\frac{M}{2}}\!\subset\!\cH_{\!M}$, and $U_{\!M}$ is a unitary parametrized by the low-pass filter $h_{n}$ which coincides with the wavelet disentangler in \cite{EvenblyEntanglementRenormalizationAnd, EvenblyRepresentationAndDesign} (see the appendix for further details). Fig.~\ref{fig:tmoar} illustrates how \eqref{eq:rgflow} can be interpreted in terms of TNR which is dual to the construction of a MERA as we will further explain in \cite{OsborneOARTNR}.
\begin{figure}[ht]
\centering
	\begin{tikzpicture}
		\draw (-1.75,0.25) node{$\rho^{(0)}_{\!M1}$} (-1.25,0.25) node{$=$} (-0.75,0.25) node{$\tfrac{1}{Z_{\!M1}}$};
		\foreach \i in {0,0.75,1.5,2.75,3.5,4.25}
		{
		\foreach \j in {-0.75,0}
		{
		\draw (0.25+\i,0.5+\j) to (0.25+\i,0.75+\j);
		}
		\draw (0.5+\i,0.25) to (0.75+\i,0.25);
		\draw (0+\i,0) to (0.5+\i,0) to (0.5+\i,0.5) to (0+\i,0.5) to (0+\i,0);
		\draw (0.25+\i,0.25) node{$A$};
		}
		\draw (-0.25,0.25) to (0,0.25);
		\draw[dotted] (2.25,0.25) to (2.5,0.25);
		\draw (2.5,0.25) to (2.75,0.25);
		\draw[dashed] (-0.25,0.25) to[out=180,in=180] (-0.25,-0.125) to (5,-0.125) to[out=0,in=0] (5,0.25);
		
		\draw[|->] (2.375,-0.5) to (2.375,-1);
		\draw (2.375, -0.75) node[right]{$\cE$};
		
		\draw (-1.75,-1.75-1) node{$\rho^{(1)}_{\!\frac{M}{2}1}$} (-1.25,-1.75-1) node{$=$} (-0.75,-1.75-1) node{$\tfrac{1}{Z_{\!M1}}$};
		\foreach \i in {0,0.75,1.5,2.75,3.5,4.25}
		{
		\foreach \j in {-0.75,0}
		{
		\draw (0.25+\i,-1.5-1+\j) to (0.25+\i,-1.25-1+\j);
		}
		\draw (0.5+\i,-1.75-1) to (0.75+\i,-1.75-1);
		\draw (0+\i,-1.5-1) to (0.5+\i,-1.5-1) to (0.5+\i,-2-1) to (0+\i,-2-1) to (0+\i,-1.5-1);
		\draw (0.25+\i,-1.75-1) node{$A$};
		}
		\draw (-0.25,-1.75-1) to (0,-1.75-1);
		\draw[dotted] (2.25,-1.75-1) to (2.5,-1.75-1);
		\draw (2.5,-1.75-1) to (2.75,-1.75-1);
		\draw[dashed] (-0.25,-1.75-1) to[out=180,in=180] (-0.25,-2.125-1) to (5,-2.125-1) to[out=0,in=0] (5,-1.75-1);
		
		\draw (0,-2.75+0.5) to (4.75,-2.75+0.5) (4.75,-2.375+0.5) to (0,-2.375+0.5);
		\draw[dotted] (4.75,-2.75+0.5) to (4.75,-2.375+0.5) (0,-2.375+0.5) to (0,-2.75+0.5);
		\draw (2.375,-2.5625+0.5) node{$U_{\!M}^{_{*}}$};
		
		\draw (0,-3.75+0.5) to (4.75,-3.75+0.5) (4.75,-4.125+0.5) to (0,-4.125+0.5);
		\draw[dotted] (4.75,-3.75+0.5) to (4.75,-4.125+0.5) (0,-4.125+0.5) to (0,-3.75+0.5);
		\draw (2.375,-3.9375+0.5) node{$U_{\!M}$};
		
		\foreach \i in {0,0.75,1.5,2.75,3.5,4.25}
		{
		\foreach \j in {-2,0}
		{
		\draw (0.25+\i,-2.375+\j+0.5) to (0.25+\i,-2.125+\j+0.5);
		}
		}
		\foreach \i in {0,2.75}
		{
		\draw (0+\i,-2.125+0.5) to (1.25+\i,-2.125+0.5);
		\draw (0+\i,-2.125+0.5) to[out=90,in=180] (0.625+\i,-1.875+0.5) to[out=0,in=90] (1.25+\i,-2.125+0.5);
		\draw (0.625+\i,-1.875+0.5) to (0.625+\i,-1.625+0.5);
		\draw (0.625+\i,-2+0.5) node{$\ptr$};
		\draw (0+\i,-4.375+0.5) to (1.25+\i,-4.375+0.5);
		\draw (0+\i,-4.375+0.5) to[out=270,in=180] (0.625+\i,-4.625+0.5) to[out=0,in=270] (1.25+\i,-4.375+0.5);
		\draw (0.625+\i,-4.625+0.5) to (0.625+\i,-4.875+0.5);
		\draw (0.625+\i,-4.5+0.5) node{$\ptr$};
		
		\draw (1.5+\i,-2.125+0.5) to (2.125+\i,-2.125+0.5);
		\draw[dotted] (2.125+\i,-2.125+0.5) to (2.125+\i,-1.875+0.5);
		\draw (1.5+\i,-2.125+0.5) to[out=90,in=180] (2.125+\i,-1.875+0.5);
		\draw (2.125+\i,-1.875+0.5) to (2.125+\i,-1.625+0.5);
		\draw (1.5+\i,-4.375+0.5) to (2.125+\i,-4.375+0.5);
		\draw[dotted] (2.125+\i,-4.375+0.5) to (2.125+\i,-4.625+0.5);
		\draw (1.5+\i,-4.375+0.5) to[out=270,in=180] (2.125+\i,-4.625+0.5);
		\draw (2.125+\i,-4.625+0.5) to (2.125+\i,-4.875+0.5);
		}
		
		\draw (-1.25,-5.5) node{$=$} (-0.75,-5.5) node{$\tfrac{1}{Z_{\!M1}}$};
		
		\foreach \i in {0,1.5,2.75,4.25}
		{
		\foreach \j in {-1.25,0}
		{
		\draw (0.625+\i,-5+\j) to (0.625+\i,-4.75+\j);
		}
		\draw (0.125+\i,-5) to (1.125+\i,-5) to (1.125+\i,-6) to (0.125+\i,-6) to (0.125+\i,-5);
		\foreach \i in {0,2.75,4.25}
		{
		\draw[thick] (1.125+\i,-5.5) to (1.625+\i,-5.5);
		}
		\draw (0.625+\i,-5.5) node{$A^{(1)}$};
		}
		\draw[thick] (-0.25,-5.5) to (0.125,-5.5);
		\draw[thick,dotted] (2.64,-5.5) to (2.825,-5.5);
		\draw[thick] (2.625,-5.5) to (2.675,-5.5);
		\draw[thick] (2.825,-5.5) to (2.875,-5.5);
		\draw[thick,dashed] (-0.25,-5.5) to[out=180,in=180] (-0.25,-6.126) to (5.875,-6.125) to[out=0,in=0] (5.875,-5.5);
	\end{tikzpicture}
	\caption{Illustration of a single renormalization group step in OAR applied to the transfer matrix $V_{\!M}$ (indices are suppressed, note that $Z_{\!M1}\!=\!Z^{(1)}_{\!\frac{M}{2}1}$). The disentangler $U_{\!M}$ can be decomposed into $2$-local operations which defines the renormalized tensor $A^{(1)}$ with increased horizontal bond dimension by the results of \cite{EvenblyRepresentationAndDesign} combined with singular value decomposition.}
	\label{fig:tmoar}
\end{figure}
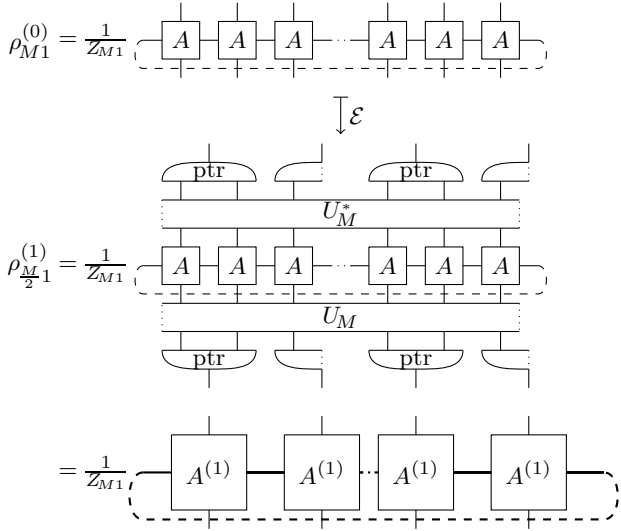

\paragraph{Infinite volume formulation.} We can avoid additional complications in the discussion of the renormalization group fixed point due to boundary conditions, necessary for the algebras $\cP_{\!M}$, $\fA_{\!M}$ at finite $M$ and $N$ by passing to an infinite volume formulation, i.e.~$M,N\!\rightarrow\!\infty$: First, we observe that imposing the asymptotic scaling conditions,
\begin{align}
\label{eq:rgcondK}
K_{1} & \!\sim\!\beta t^{(3)}N^{-1}, & K^{*}_{2} & \!\sim\!\beta t^{(1)}N^{-1},
\end{align}
for $\beta, t^{(3)}, t^{(1)}\!>\!0$ for $N\!\rightarrow\!\infty$, provides a Gibbs state, $\rho_{MN}\!\stackrel{N\rightarrow\infty}{\rightarrow}\!\rho_{M}\!=\!\tfrac{1}{Z_{M}}e^{-\beta H_{\!M}}$, of the transverse-field Ising Hamiltonian at inverse temperature $\beta$ \cite{SuzukiQuantumIsingPhases} as a consequence of Trotter's product formula \cite{ReedMethodsOfModern1},
\begin{align}
\label{eq:2diH}
H_{\!M} & \!=\! -\!\!\sum_{j=-M}^{M-1}\!\!\big(t^{(3)}\sigma^{(3)}_{j}\sigma^{(3)}_{j+1}\!+\!t^{(1)}\sigma^{(1)}_{j}\big).
\end{align}
Second, we note that the definition of the dual quantum channels $\alpha:\fA_{M}\!\rightarrow\!\fA_{2M}$ is compatible with taking the infinite volume limit, $\varinjlim_{M}\fA_{M}\!=\!\cup_{M}\fA_{M}\!=\!\fA$, in the sense of quasi-local algebras \cite{BratteliOperatorAlgebrasAnd2}, which leads to a description of the limit $M\!\rightarrow\!\infty$ in terms of the fermion algebra, $\fA\!=\!\fA_{\CAR}(\fh)$, with one-particle space $\fh\!=\!\ltwo(\Z)$, and the renormalization group transformation, $\alpha:\fA\!\rightarrow\!\fA$, defined by the analogue of \eqref{eq:rg1p}. The dynamics on $\fA$ is determined by the Hamiltonian $H$, formally given by \eqref{eq:2diH} for $M\!\rightarrow\!\infty$, which is still well-defined as a derivation on strictly local elements of $\fA$ \cite{BratteliOperatorAlgebrasAnd2}. In this limit, the Gibbs states $\rho_{\!M}$ provide quasi-free KMS-states $\omega_{\!\beta}:\fA\!\rightarrow\!\C$ determined by the two-point function:
\begin{align}
\label{eq:KMS2p}
\omega_{\!\beta}(\!a(\xi)a^{\!\dag}\!(\eta)\!) & \!\!=\!\! \langle\xi,\!C^{(1)}_{\beta}\!\eta\rangle, & \!\!\!\!\!\omega_{\!\beta}(\!a^{\!\dag}\!(\xi)a^{\!\dag}\!(\eta)\!) & \!\!=\!\!\langle\bar{\xi},\!C^{(2)}_{\beta}\!\eta\rangle.
\end{align}
The covariance operators $C^{(1)}_{\beta}$, $C^{(2)}_{\beta}$ have momentum-space kernels,
\begin{align}
\label{eq:KMScov}
C^{(1)}_{\beta}\!(\theta,\theta') & \!=\! \pi\big(1\!+\!\Re\big(\!\tfrac{z_{\theta}}{|z_{\theta}|}\!\big)\!\tanh(\beta|z_{\theta}|)\!\big)\delta(\theta\!-\!\theta'), \\ \nonumber
C^{(2)}_{\beta}\!(\theta,\theta') & \!=\! -i\pi\Im\big(\!\tfrac{z_{\theta}}{|z_{\theta}|}\!\big)\!\tanh(\beta|z_{\theta}|)\delta(\theta\!-\!\theta'),
\end{align}
where $z_{\theta}\!=\!t^{(1)}-e^{i\theta}t^{(3)}$. The expressions remain meaningful in the limit $\beta\!\rightarrow\!\infty$ providing a ground state of $H$ on $\fA$. Evaluating the renormalization group flow \eqref{eq:rgflow} results in sequences of renormalized states $\omega^{(m)}_{\beta}\!=\!\omega_{\beta}\!\circ\!\alpha^{m}$ which are quasi-free by construction and, thus, determined by their two-point functions:
\begin{widetext}
\begin{align}
\label{eq:KMS2prg}
\omega^{(m)}_{\beta}(a(\xi)a^{\dag}(\eta)\!) & \!=\!\langle R^{m}(\xi),C^{(1)}_{\beta}R^{m}(\eta)\rangle \!=\! \tfrac{1}{4\pi}\!\int^{\pi}_{-\pi}\!\!d\theta\!\ \big(1\!+\!\Re\big(\!\tfrac{z_{\theta}}{|z_{\theta}|}\!\big)\!\tanh(\beta|z_{\theta}|)\!\big)2^{m}\Big(\prod_{n=0}^{m-1}|m_{0}(2^{n}\theta)|^{2}\Big) \overline{\hat{\xi}_{2^{m}\theta}}\hat{\eta}_{2^{m}\theta}, \\ \nonumber
\omega^{(m)}_{\beta}(a^{\dag}(\xi)a^{\dag}(\eta)\!) & \!=\!\langle R^{m}(\bar{\xi}),C^{(2)}_{\beta}R^{m}(\eta)\rangle \!=\! -\tfrac{i}{4\pi}\!\int^{\pi}_{-\pi}\!\!d\theta\!\ \Im\big(\!\tfrac{z_{\theta}}{|z_{\theta}|}\!\big)\!\tanh(\beta|z_{\theta}|)2^{m}\Big(\prod_{n=0}^{m-1}|m_{0}(2^{n}\theta)|^{2}\Big) \hat{\xi}_{-2^{m}\theta}\hat{\eta}_{2^{m}\theta}.
\end{align} 
\end{widetext}
Fixed points and admissible scaling limits are determined by analyzing the convergence of \eqref{eq:KMS2prg} for $m\!\rightarrow\!\infty$ under suitable renormalization conditions imposed on the couplings $t^{(1)}, t^{(3)}$ and the inverse temperature $\beta$.

\paragraph{The fixed point at criticality.} In the quantum spin-chain formulation, the critical line corresponds to equal couplings $t^{(3)}\!=\!t^{(1)}\!=\!t$ in the Hamiltonian \eqref{eq:2diH} in the limit $\beta\!\rightarrow\!\infty$, which is equivalent to $K_{1}\!\approx\!K^{*}_{2}$ (at large $N\!\gg\!1$ by \eqref{eq:rgcondK}), i.e.~the critical line of the two-dimensional Ising model given by the tensor \eqref{eq:2dit} corresponding to the well-know critical coupling $K\!=\!K_{1}\!=\!K_{2}\!=\!\tfrac{1}{2}\ln(1\!+\!\sqrt{2})$ in the isotropic case. In view of \eqref{eq:KMS2prg}, we have $z_{\theta}\!=\!t(1-e^{i\theta})$ and $|z_{\theta}|^{2}\!=\!4t^{2}\sin(\tfrac{1}{2}\theta)^{2}$. Using the change of variables $k\!=\!2^{m}\theta$ and noting that $\tanh(\beta|z_{\theta}|)\!\stackrel{\beta\rightarrow\infty}{\rightarrow}\!1\!-\!\delta_{\theta,0}$, we find:
\begin{widetext}
\begin{align}
\label{eq:crit2prg}
\omega^{(m)}_{\beta=\infty}(a(\xi)a^{\dag}(\eta)\!) & \!=\! \tfrac{1}{4\pi}\!\int^{2^{m}\pi}_{-2^{m}\pi}\!\!dk\!\ \big(1\!+\!\tfrac{1-\cos(2^{-m}k)}{2|\sin(\frac{1}{2}2^{-m}k)|}\!\big)\Big(\prod_{n=1}^{m}|m_{0}(2^{-n}k)|^{2}\Big) \overline{\hat{\xi}_{k}}\hat{\eta}_{k} \!\stackrel{m\rightarrow\infty}{\rightarrow}\! \tfrac{1}{4\pi}\!\int^{\infty}_{-\infty}\!\!dk\!\ |\hat{s}(k)|^{2} \overline{\hat{\xi}_{k}}\hat{\eta}_{k}, \\ \nonumber
\omega^{(m)}_{\beta=\infty}(a^{\dag}(\xi)a^{\dag}(\eta)\!) & \!=\! \tfrac{i}{4\pi}\!\int^{2^{m}\pi}_{-2^{m}\pi}\!\!dk\!\ \tfrac{\sin(2^{-m}k)}{2|\sin(\frac{1}{2}2^{-m}k)|}\Big(\prod_{n=1}^{m}|m_{0}(2^{-n}k)|^{2}\Big) \hat{\xi}_{-k}\hat{\eta}_{k} \!\stackrel{m\rightarrow\infty}{\rightarrow}\! \tfrac{i}{4\pi}\!\int^{\infty}_{-\infty}\!\!dk\!\ \sign(k)|\hat{s}(k)|^{2} \hat{\xi}_{-k}\hat{\eta}_{k},
\end{align} 
\end{widetext}
by Lebesgue's dominated convergence theorem applied to $\prod_{n=1}^{m}m_{0}(2^{-n}k)\!\stackrel{m\rightarrow\infty}{\rightarrow}\!\hat{s}(k)$ \cite{DaubechiesTenLecturesOn}, see also \cite[Lem.~3.7]{OsborneCFTapprox} for an adapted decay estimate for $m_{0}$. By passing to the self-dual chiral Majorana fields, $\psi_{\pm|j}\!=\!e^{\pm i\frac{\pi}{4}}a_{j}\!+\!e^{\mp i\frac{\pi}{4}}a^{\dag}_{j}$, we recognize that the limits in \eqref{eq:crit2prg} are the vacuum two-point functions of the $c=\tfrac{1}{2}$ free-fermion conformal field theories (CFTs) of the two chiral halves of the critical Ising fixed point:
\begin{align}
\label{eq:crit2pmaj} \nonumber
\omega(\psi_{\!\pm}\!(\xi\!\ast\!s)\psi_{\!\pm}\!(\eta\!\ast\!s)^{\!*}\!) & \!=\!\tfrac{1}{\pi}\!\!\!\int^{\infty}_{-\infty}\!\!\!\!\!\!dk\!\ \tfrac{1}{2}(1\!\pm\!\sign(k)\!)|\hat{s}(k)|^{2} \overline{\hat{\xi}_{k}}\hat{\eta}_{k}, \\
\omega(\psi_{\!\pm}\!(\xi\!\ast\!s)\psi_{\!\mp}\!(\eta\!\ast\!s)^{\!*}\!) & \!=\!0,
\end{align} 
where $(\xi\!\ast\!s)(x)\!=\!\sum_{j\in\Z}\xi_{j}s(x\!-\!j)$ for $\xi\in\fh$. We directly infer from \eqref{eq:crit2pmaj} that the scaling function $s$ controls the resolution at which the CFT is probed.

\paragraph{Error bounds on fermions correlations.} It is an immediate consequence of the construction that explicit error bounds on the approximation of dynamical fermionic $n$-point functions of the scaling limit state $\omega$ can be derived using the methods of \cite{OsborneCFTapprox, OsborneCFTsim}:
\begin{align}
\label{eq:ncor}
|\omega^{(\!m\!)}\!(a^{(\dag)}_{t^{(\!0\!)}_{1}}\!(\xi_{1}\!)...a^{(\dag)}_{t^{(\!0\!)}_{n}}\!(\xi_{n}\!)\!)\!-\!\omega(a^{(\dag)}_{t_{1}}\!(\xi_{1}\!)...a^{(\dag)}_{t_{n}}\!(\xi_{n}\!)\!)| & \!\leq\!\delta,
\end{align}
given a set of one-particle vectors $\xi_{1},...,\xi_{n}\!\in\!\fh$ and effective lattice times $t^{(0)}_{1},\!...,t^{(0)}_{n}$ as well as continuum times $t_{1},...,t_{n}$ (referring to the effective dynamics $H$ after rescaling $m$-times and the massless free-fermion dynamics in the scaling limit respectively). In particular, we find,
\begin{align}
\label{eq:error}
\delta & \!=\!\delta(m,T)\!\leq\!2^{-m}C_{T},
\end{align}
for $|t_{1}|,...,|t_{n}|\!\in\![0,T]$, large effective lattice times $t^{(0)}_{i}\!\sim\!2^{m}t_{i}$, and some constant $C_{T}\!>\!0$ otherwise only depending on one-particle norms of $\xi_{1},\!...,\xi_{n}$ and the scaling function $s$. In general, the error in \eqref{eq:ncor} can only be small for large effective lattice times just as the equal-time correlation approximate their continuum counterparts at large distances, as seen from \eqref{eq:crit2pmaj}. But, here an exponential separation of effective lattice and continuum times is not necessary at the expense of a slower decay of the error $\delta$ (see appendix).
\paragraph{Instability of the fixed point at criticality.} The question of stability of fixed points in the framework of TNR has been of interest recently \cite{KennedyTensorRGApproach, KennedyTensorRenormalizationGroup}. Although, we cannot address this question for OAR in full detail in this Letter, we can make the following observation: In the space of quasi-free (initial) states characterized by covariance operators $C\!=\!C_{\beta}(t^{(1)}, t^{(3)})$ in the sense of \eqref{eq:KMS2p}, it is an immediate consequence of \eqref{eq:KMS2prg} that the critical state given by $C_{\textup{crit.}}\!=\!C_{\beta=\infty}(t,t)$ is unstable, because $\tfrac{z_{\theta}}{|z_{\theta}|}\!\stackrel{m\rightarrow\infty}{\rightarrow}\!\sign(\lambda)$ and $|z_{\theta}|\!\stackrel{m\rightarrow\infty}{\rightarrow}\!t^{(1)}|\lambda|$ for $\theta\!=\!2^{-m}k$ and $\lambda\!=\!1\!-\!\tfrac{t^{(3)}}{t^{(1)}}\!\in\![-\infty,0)\!\cup\!(0,1]$ (non-critical). In particular, at $\beta\!=\!\infty$, the states are driven towards: (1) the disorder fixed point $\lambda\!=\!1$ ($t^{(1)}\!=\!\textup{const.}, t^{(3)}\!=\!0$) for $\lambda\in(0,1]$, or (2) the order fixed point $\lambda\!=\!-\infty$ ($t^{(1)}\!=\!0, t^{(3)}\!=\!\textup{const.}$) for $\lambda\!\in\![-\infty,0)$. By a similar reasoning that led to \eqref{eq:crit2prg}, the disorder fixed point is given by the Fock state with respect to $a,a^{\dag}$ while the order fixed point is given by the anti-Fock state (resulting from an equal weight mixture of the two extremal ground states of $H_{\!M}$ at $t^{(1)}\!=\!0$).

\paragraph{Spin-spin correlations.}
The correspondence between quasi-free states on $\fA$ and even states on the infinite-volume Pauli algebra $\cP\!=\!\otimes_{j\in\Z}M_{2}(\C)$ \cite{EvansQuantumSymmetriesOn} allows for a characterization of the critical fixed point in terms of spin-spin correlation functions,
\begin{align}
\label{eq:spinspin} \nonumber
\omega(\sigma^{(3)}_{j_{1}}...\sigma^{(3)}_{j_{2n}}) &\!=\!\omega\Big(\prod_{k=1}^{n}\prod_{ l_{k}=j_{2k-1}}^{j_{2k}-1}\!\!\!\!\!\Psi(0,i\delta_{l_{k}})\Psi(\delta_{l_{k}\!+\!1},0)\!\Big), \\
\omega(\sigma^{(3)}_{j_{1}}...\sigma^{(3)}_{j_{2n+1}}) & \!=\!0,\ j_{1} \!\leq\!...\!\leq\!j_{2n}\!\leq\!j_{2n+1},
\end{align}
where $\Psi(\xi,\eta)\!=\!a(\xi\!-\!i\eta)\!+\!a^{\dag}(\overline{\xi\!+\!i\eta})$, $\xi,\eta\!\in\!\ltwo(\Z)$, is Araki's self-dual field \cite{ArakiOnTheXY}. These correlation functions are precisely the scaling limits of the Ising correlation functions \eqref{eq:2dicor} at criticality $K_{1}\!=\!K_{2}^{*}$ for $k_{1},...,k_{n}\!=\!0$. The quasi-free structure of $\omega$ allows for the evaluation of \eqref{eq:spinspin} in terms of a Pfaffian \cite{EvansQuantumSymmetriesOn}, which further reduces to well-known Toeplitz determinant \cite{MontrollCorrelationsAndSpontaneous, SchultzTwoDimensionalIsing} with the crucial difference that scaling-limit two-point function is given by \eqref{eq:crit2prg}. The real-time, analytic continuations of the critical Ising correlation functions with $k_{1},...,k_{n}\!\neq\!0$ can be obtained from \eqref{eq:crit2prg} by means of the scaling limit of the time-evolution of $H_{\!M}$ (see the appendix for a sketch).

\paragraph{Other scaling limits.} Inspecting \eqref{eq:KMS2prg} it is straightforward to construct massive and finite-temperature scaling limits: If we impose the renormalization conditions $\lambda\!=\!1\!-\!\tfrac{t^{(3)}}{t^{(1)}}\!\sim\!2^{-m}\mu_{0}\!>\!$ and $\beta\!\sim\!2^{m}\beta_{0}$ for arbitrary $\mu_{0}\!\geq\!0$ and $\beta\!>\!0$, we will obtain the equilibrium state at temperature $\beta_{0}$ of a free fermion quantum field of mass $m_{0}$:
\begin{widetext}
\begin{align}
\label{eq:KMS2pcont}
\omega_{\mu_{0},\beta_{0}}\!(a(\xi\!\ast\!s)a^{\!\dag}(\eta\!\ast\!s)\!) & \!=\!\tfrac{1}{4\pi}\!\!\int^{\infty}_{-\infty}\!\!\!dk\!\ \big(1\!+\!\tfrac{\mu_{0}}{\omega_{\mu_{0}}\!(k)}\!\tanh(\beta_{0}t\omega_{\mu_{0}}(k)\!)\!\big)|\hat{s}(k)|^{2} \overline{\hat{\xi}_{k}}\hat{\eta}_{k}, \\ \nonumber
\omega_{\mu_{0},\beta_{0}}\!(a^{\!\dag}(\xi\!\ast\!s)a^{\!\dag}(\eta\!\ast\!s)\!) & \!=\!\tfrac{i}{4\pi}\!\!\int^{\infty}_{-\infty}\!\!\!dk\!\ \tfrac{k}{\omega_{\mu_{0}}\!(k)}\tanh(\beta_{0}t\omega_{\mu_{0}}(k)\!)|\hat{s}(k)|^{2} \hat{\xi}_{-k}\hat{\eta}_{k}
\end{align} 
\end{widetext}
where $\omega_{\mu_{0}}\!(k)^{2}\!=\!\mu_{0}^{2}\!+\!k^{2}$ is the massive continuum dispersion relation, $(\xi\!\ast\!s)(x)\!=\!\sum_{j\in\Z}\xi_{j}s(x\!-\!j)$ for $\xi\in\fh$, and $t^{(3)}\!\stackrel{m\rightarrow\infty}{\rightarrow}\!t$. As before, the scaling function $s$ controls the resolution at which the continuum quantum field is probed.

\paragraph{Discussion.}
We have presented an explicit description of the critical fixed point of the two-dimensional classical Ising model using OAR which may be understood as a Wilson-Kadanoff RG scheme dual to tensor-network methods. In particular, if OAR is applied to density matrices given in terms of transfer matrices of classical lattice systems, it is operationally dual to a (thermal) MERA derived from TNR \cite{EvenblyTNRMERA}. Our explicit representation of the critical fixed point relies on an implementation of OAR using wavelet methods that was previously introduced in \cite{StottmeisterOperatorAlgebraicRenormalization, MorinelliScalingLimitsOf, OsborneCFTapprox, OsborneCFTsim}, and the duality with TNR is manifestly exhibited by the unitary defining the coarse-graining channel $\cE$ (see \eqref{eq:cg} in the appendix), which directly corresponds to the exact disentangler of Evenbly and White for the ground state of the Ising quantum chain \cite{EvenblyEntanglementRenormalizationAnd, EvenblyRepresentationAndDesign}. In our construction of the scaling limit, a particularly important role is played by the scaling function associated with a given low-pass filter, as this function controls the resolution at which the fixed-point tensor is probed at unit scale -- either in terms of fermionic or spin-spin correlation functions. As a direct consequence of this feature we explicitly observe a universal large-scale behavior independent of the specific choice of scaling functions. Another important advantage of our method over other approaches such as the exact MERA is the provision of explicit, provable error bounds on the approximation of correlation functions for sufficiently regular scaling functions that are independent of the design problem of Hilbert-pair wavelets \cite{HaegemanRigorousFreeFermion}. Such error bounds allow for a direct understanding of the simulation of QFTs/CFTs by quantum computers \cite{OsborneCFTsim}.
We have exhibited a direct correspondence of the critical fixed point with the vacuum (or Neveu-Schwarz) sector of the Ising CFT with an explicit formula for the two-point functions of the self-dual chiral Majorana field (see \eqref{eq:crit2pmaj}). By our method, it is possible obtain fixed points corresponding to other sectors, e.g., the Ramond sector, by working in a finite-volume setting including different, e.g., anti-periodic, boundary conditions, which will be discussed elsewhere \cite{StottmeisterQuantumScalingLimitI}. In addition, we are planning to further clarify the relation of our construction of the scaling limit of the Ising model with previously known results about the Ising QFT/CFT -- specifically via spin-spin correlation functions \cite{McCoyTwoDimensionalIsing, AbrahamNPointFunctions1, AbrahamNPointFunctions2, BarievManyPointCorrelation, AbrahamPlanarIsingFerromagnet, SatoHolonomicQuantumFieldsV} and the explicit construction of the spin field operator \cite{RuijsenaarsIntegrableQuantumField}.

\begin{acknowledgments}
This work was supported, in part, by the Quantum Valley Lower Saxony, the Deutsche Forschungsgemeinschaft (DFG, German Research Foundation) – Project-ID 274200144 – SFB 1227, and under Germanys Excellence Strategy EXC-2123 QuantumFrontiers 390837967. AS was in part supported by the MWK Lower Saxony within the Stay Inspired program.
\end{acknowledgments}

\appendix

\widetext

\section{Appendix}

\section{Coarse graining $\cE$: spin chain representation}
Let us briefly explain the structure of the coarse graining channel $\cE$ in the spin chain representation (additional details can be found in \cite{OsborneOARTNR}):
For finite-length spin chains, the Jordan-Wigner transformation,
\begin{align}
\label{eq:jw}
a_{j} & \!=\!\big(\prod_{-M\leq l<j}\sigma^{(1)}_{l}\big)\tfrac{1}{2}(\sigma^{(3)}_{j}\!+\!i\sigma^{(2)}_{j}), & j &\in\Z
\end{align}
provides a $*$-isomorphism between the Pauli algebra $\cP_{\!M}$ and the complex fermion algebra $\fA_{\!M}$. In terms of Araki's self-dual field \cite{ArakiOnTheXY}, $\Psi(\xi,\eta)\!=\!a(\xi\!-\!i\eta)\!+\!a^{\dag}(\overline{\xi\!+\!i\eta})$, the inverse of this transformation can be written as:
\begin{align}
\label{eq:jwsd}
\sigma^{(3)}_{j} & \!=\! \big(\prod_{-M\leq l<j}\!\!\!\Psi(\delta_{l},0)\Psi(0,i\delta_{l})\big)\Psi(\delta_{j},0), & \sigma^{(1)}_{j} & \!=\! \Psi(\delta_{j},0)\Psi(0,i\delta_{j}).
\end{align}
We can now exploit the Jordan-Wigner transformation to find an explicit representation of the dual renormalization group channel $\alpha$ given in \eqref{eq:rg1p}.First, we observe the associated state correspondence between the natural representations of $\cP_{\!M}$ on $\cH_{\!M}\!=\!\otimes_{j=-M}^{M-1}\C^{2}$ and $\fA_{\!M}$ on the anti-symmetric Fock space $\fF_{\!-}\!(\fh_{\!M})$:
\begin{align}
\label{eq:jws}
|j_{1},\!...,\!j_{n}\rangle & \!\!=\!\! a^{\dag}_{j_{1}}\!...a^{\dag}_{j_{n}}\!\Omega_{\!M} \!\!=\!\!(\!-\!1)^{\!|j|}\!|-\rangle_{\!-\!M}\!\!\otimes\!...\!\otimes\!|+\rangle_{\!j_{1}}\!\!\otimes\!...\!\otimes\!|+\rangle_{\!j_{n}}\!\!\otimes\!...\!\otimes\!|-\rangle_{\!M\!-\!1} \!\!=\!\! (\!-\!1)^{\!|j|}\!|-_{\!-\!M},\!...,\!+_{j_{1}},\!...,\!+_{j_{n}},\!...,\!-_{\!M\!-\!1}\rangle, \\ \nonumber
\end{align}
for $-M\!\leq\!j_{1}\!<\!...\!<j_{n}\!\leq\!M\!-\!1$, where $\Omega_{M}\!=\!\otimes_{j=-M}^{M-1}|-\rangle\!=\!|-,...,-\rangle$ is the Fock vacuum, $|j|\!=\!(\!-\!1)^{\sum^{n}_{i=1}j_{i}}$, and $|\pm\rangle$ are the $\pm1$-eigenstates of $\sigma^{(1)}$. Second, we decompose the one-particle isometry $R\!=\!u_{2M}\!\circ\!\iota$ given in \eqref{eq:rg1p} into a unitary $u_{2M}$ on $\fh_{2M}$ and the trivial embedding $\iota(\fh_{\!M})\!\subset\!\fh_{2M}$ given by the identification of the scaled lattice $2\Lambda_{\!M}$ as sublattice of $\Lambda_{2M}$. Clearly, the unitary is not uniquely fixed on the orthogonal complement of $\iota(\fh_{\!M})$, but there is a natural choice associated with a length-$2N$ low-pass filter $\{h_{n}\}$ given by a corresponding high-pass filter $\{g_{n}\!=\!(-1)^{n}\bar{h}_{-n+1+2N}\}$ \cite{DaubechiesTenLecturesOn}:
\begin{align}
\label{eq:disentangler}
(u_{2M}\xi)_{j} & \!=\! \sum_{l\in2\Lambda_{\!M}}\xi_{l}\sum_{n\in\Z}h_{n}\delta_{l+n,j}+\sum_{l'\in\Lambda_{2M}\setminus 2\Lambda_{\!M}}\xi_{l'}\sum_{n\in\Z}g_{n+1}\delta_{l'+n,j}, & \xi & \in\fh_{2M}.
\end{align}
This choice precisely corresponds to the representation of the discrete wavelet transform (on $L^{2}(S^{1})$ or $L^{2}(\R)$) as a unitary circuit by Evenbly-White \cite{EvenblyRepresentationAndDesign}. We note that by construction the unitary $U_{2M}\!=\!\Gamma(u_{2M})$ in Fig.~\ref{fig:tmoar} acting on $\cH_{2M}\!=\!\fF_{\!-}\!(\fh_{2M})$ is given by the multiplicative second quantization of $u_{\!M}$, while the unital $*$-morphism $\Gamma(\iota)$ is dual to the partial trace $\ptr\!:\!\cH_{2M}\!\rightarrow\!\cH_{\!M}$. In the limit $M\!\rightarrow\!\infty$, \eqref{eq:disentangler} is still meaningful if the low-pass filter is considered in non-periodic form as arising from a scaling function $s\in\C^{r}(\R)$. Similarly, the partial trace is performed with respect to every other site of the infinite-length spin chain. In summary we have,
\begin{align}
\label{eq:cg}
\cE & \!=\! \ptr\circ\Ad_{U},
\end{align}
and $U$ has the following matrix elements with respect to the $\sigma^{(1)}$-basis:
\begin{align}
\label{eq:disentanglerbasis}
\langle-...\!+_{j'_{1}}\!...\!+_{j'_{n}}\!...\!-\!|U|\!-\!...\!+_{j_{1}}\!...\!+_{j_{m}}\!...\!-\rangle & \!=\! \delta_{n,m}(-1)^{|j'|+|j|}\sum_{\{n\}}\prod_{l=1}^{n}c^{(j_{l})}_{n_{l}}\delta_{j'_{l},j_{l}+n_{l}},
\end{align}
where $c^{(j)}_{n}\!=\!\left\{\begin{matrix} h_{n} & : & j\in2\Z \\ g_{n+1} & : & j\in\Z\setminus2\Z \end{matrix}\right.$.

\section{Even states: spin-spin correlation functions}
We provide additional details on the computation of spin-spin correlation functions and the definition of dynamics on the Pauli algebra $\cP$ following \cite{SchultzTwoDimensionalIsing, ArakiOnTheXY}. The Pauli algebra $\cP$ and the fermion algebra $\fA$ are both $\Z_{2}$-graded in a compatible way, i.e.~we have order-$2$ automorphisms $\Theta$ of $\cP$ and $\fA$ (denoted by the same symbol) given by:
\begin{align}
\label{eq:grad}
\Theta(\sigma^{(1)}_{j}) &\!=\!\sigma^{(1)}_{j}, & \Theta(\sigma^{(3)}_{j}) &\!=\!-\sigma^{(3)}_{j}, \\ \nonumber
\Theta(a(\xi)) & \!=\! -a(\xi), & \xi & \in\fh.
\end{align}
It is known that there is a bijective correspondence between even states, those invariant under $\Theta$, on $\cP$ and $\fA$, and it is easy to see that the Gibbs states $\omega_{\beta}$ as well as the ground state $\omega_{\infty}$ are even states on $\fA$ \cite{EvansQuantumSymmetriesOn}. While the Jordan-Wigner transformation \eqref{eq:jw} does not extend to a $*$-isomorphism between the infinite-length Pauli algebra $\cP$ and the complex fermion algebra $\fA$, it does so for the even parts $\cP^{(0)}$ and $\fA^{(0)}$ with respect to the $\Z_{2}$-grading $\Theta$. In particular, we have:
\begin{align}
\label{eq:2ptjw}
\sigma^{(3)}_{j}\sigma^{(3)}_{j'} & \!=\! (a_{j}\!+\!a^{\dag}_{j})\big(\prod_{j\leq l<j'}\!\!\![a^{\dag}_{l},a_{l}]\big)(a_{j'}\!+\!a^{\dag}_{j'}) \!=\! \prod_{j\leq l<j'}\!\!\!(a_{l}\!-\!a^{\dag}_{l})(a_{l+1}\!+\!a^{\dag}_{l+1}) \!=\! \prod_{j\leq l<j'}\!\!\!\Psi(0,i\delta_{l})\Psi(\delta_{l+1},0), & j & \leq j', 
\end{align}
where $\Psi(\xi,\eta)\!=\!a(\xi\!-\!i\eta)\!+\!a^{\dag}(\overline{\xi\!+\!i\eta})$ is Araki's self-dual field. This allows for a direct evaluation of even quasi-free states, e.g., $\omega_{\beta}$ and $\omega$, on $\cP$ in terms of $\fA$: By means of \eqref{eq:2ptjw} we obtain the following expression for $n$-point spin-spin correlation functions (cp.~\eqref{eq:spinspin}):
\begin{align}
\label{eq:spinspinex}
\omega(\sigma^{(3)}_{j_{1}}...\sigma^{(3)}_{j_{2n}}) & \!=\! \omega\Big(\prod_{j_{1}\leq l<j_{2}}\!\!\!\Psi(0,i\delta_{l_{1}})\Psi(\delta_{l_{1}+1},0)...\prod_{j_{2n-1}\leq l<j_{2n}}\!\!\!\Psi(0,i\delta_{l_{n}})\Psi(\delta_{l_{n}+1},0)\Big), &  &  \\ \nonumber
\omega(\sigma^{(3)}_{j_{1}}...\sigma^{(3)}_{j_{2n+1}}) & \!=\! 0, & j_{1} & \!\leq\!...\!\leq\!j_{2n}\!\leq\!j_{2n+1},
\end{align}
which can be evaluated for quasi-free state on $\fA$ as these are determined by their two-point function \cite{EvansQuantumSymmetriesOn} (odd correlators vanish identically):
\begin{align}
\label{eq:pfaffian}
\omega(\Psi(\xi_{1},\!\eta_{1})...\Psi(\xi_{2n},\!\eta_{2n})\!) & \!=\! \textup{Pf}(\!(\omega(\Psi(\xi_{i},\!\eta_{i})\Psi(\xi_{j},\!\eta_{j})\!)\!)_{ij}) = \!\!\!\!\!\!\sum_{\substack{J,K\in D_{I_{2n}}, \\ J\sqcup K = I_{2n}, \\ |J|=|K|, J<K}}\!\!\!\!\!\!\!\!(-1)^{\binom{n}{2}}\vep(J,\!K)\!\prod_{i=1}^{n}\omega(\Psi(\xi_{j_{i}},\!\eta_{j_{i}})\Psi(\xi_{k_{i}},\!\eta_{k_{i}})\!).
\end{align}
Here, $I_{2n}\!=\!\{1,...,2n\}$ and $D_{I_{2n}}$ denotes the ordered subsets of $I_{2n}$. In particular, the sum runs over disjoined partitions of $I_{2n}$ into equally sized ordered subsets $J\!=\!\{j_{1},...,j_{n}\}$, $K\!=\!\{k_{1},...,k_{n}\}$ such that $j_{i}\!<\!k_{i}$, $i\!=\!1,...,n$. The sign $\vep(J,K)$ is given by the signature of the permutation $\binom{I}{JK}\!=\!\binom{1,...,2n}{j_{1},...,j_{n},k_{1},...,k_{n}}$. For the infinite-length transverse-field Ising spin chain at criticality, the relevant two-point functions of the self-dual field are determined by \eqref{eq:crit2prg}:
\begin{align}
\label{eq:2ptsd}
\omega(\Psi(0,i\delta_{j})\Psi(\delta_{j'},0)\!) & \!=\! 2\Re\big(\omega(a_{j}a^{\dag}_{j'})\!-\!\omega(a^{\dag}_{j}a^{\dag}_{j'})\!\big) \!=\! \tfrac{1}{2\pi}\int dk\!\ |\hat{s}(k)|^{2}\big(\!\cos(\!(j\!-\!j')k)\!+\!\sign(k)\sin(\!(j\!-\!j')k)\!\big), \\ \nonumber
\omega(\Psi(0,i\delta_{j})\Psi(0,i\delta_{j'})\!) & \!=\! -2i\Im\big(\omega(a_{j}a^{\dag}_{j'})\!-\!\omega(a^{\dag}_{j}a^{\dag}_{j'})\!\big) \!=\! 0, \\ \nonumber
\omega(\Psi(\delta_{j},0)\Psi(\delta_{j'},0)\!) & \!=\! 2i\Im\big(\omega(a_{j}a^{\dag}_{j'})\!+\!\omega(a^{\dag}_{j}a^{\dag}_{j'})\!\big) \!=\! 0.
\end{align}
As only the mixed correlations between $\Psi(0,i\delta_{j})$ and $\Psi(\delta_{j'},0)$ are non-vanishing, the Pfaffian in \eqref{eq:pfaffian} can be further reduced. Specifically, the two-point spin-spin correlation function can, thus, be evaluated in terms of a Toeplitz determinant \cite{MontrollCorrelationsAndSpontaneous, SuzukiQuantumIsingPhases}:
\begin{align}
\label{eq:toeplitz}
\omega(\sigma^{(3)}_{j}\sigma^{(3)}_{j'}) & \!=\! \det\begin{pmatrix} C^{(3)}_{-1} & C^{(3)}_{-2} & ... & C^{(3)}_{j-j'} \\ C^{(3)}_{0} & C^{(3)}_{-1} & ... & C^{(3)}_{j-j'+1} \\ ... & C^{(3)}_{0} & ... & ... \\ C^{(3)}_{j'-j-2} & ... & ... & C^{(3)}_{-1} \end{pmatrix},
\end{align}
where $C^{(3)}_{j-j'}\!=\!\omega(\Psi(0,i\delta_{j})\Psi(\delta_{j'},0)\!)\!=\!-\omega(\Psi(\delta_{j},0)\Psi(0,i\delta_{j'})\!)\!=\!C^{(3)}_{j'-j}$.

The evaluation of dynamical spin-spin correlation functions corresponding to real-time analytic continuations of the spin-spin correlators of the two-dimensional Ising model in infinite-volume \eqref{eq:2dicor} requires additional effort. To this end, following \cite{ArakiOnTheXY} we introduce \emph{tail} and \emph{string operators},
\begin{align}
\label{eq:tail}
T_{\!M} & \!=\! \prod_{-M\leq l\leq 0}\sigma^{(1)}_{l} \!=\! \prod_{-M\leq l\leq 0}\!\!\![a^{\dag}_{l},a_{l}] \!=\! \prod_{-M\leq l\leq 0}\!\!\!\Psi(\delta_{l},0)\Psi(0,i\delta_{l}), \\ \nonumber
S_{j} & \!=\! \left\{\begin{matrix} \prod_{0<l<j}[a^{\dag}_{l},a_{l}] & : & j>0 \\ \prod_{j\leq l\leq 0}[a^{\dag}_{l},a_{l}] & : & j\leq 0 \end{matrix}\right.,
\end{align}
which entails,
\begin{align}
\label{eq:jwtail}
\sigma^{(3)}_{j} & \!=\! T_{\!M}S_{j}(a_{j}\!+\!a^{\dag}_{j}),
\end{align}
to understand the relation between $\cP_{\!M}$ and $\fA_{\!M}$ via the Jordan-Wigner transformation \eqref{eq:jw} in the infinite-length limit $M\!\rightarrow\!\infty$ (see also \cite[Ex.~6.2.14]{BratteliOperatorAlgebrasAnd2}). Although, $T_{\!M}$ does not exist in this limit, the induced order-$2$ quasi-free automorphism by its adjoint action $\Theta_{M|-}(a_{j})\!=\!\Ad_{T_{\!M}}(a_{j})\!=\!\sign(j)a_{j}$, where $\sign(j)\!=\!\left\{\begin{matrix} +1 & : & 0\!<\!j\leq\!M\!-\!1 \\ -1 & : & -M\!\leq\!j\!\leq\!0 \end{matrix}\right.$ remains well-defined (denoted by $\Theta_{-}$). Therefore, $\cP$ and $\fA$ can be realized as subalgebras of the crossed product $\hat{\fA}\!=\!\fA\!\rtimes_{\theta_{-}}\!\Z_{2}$, i.e., the algebra generated by $\fA$ and a self-adjoint unitary $T$ such that:
\begin{align}
\label{eq:tailinf}
\Theta_{-}(a(\xi)\!) & \!=\! Ta(\xi)T, & \xi & \in\fh.
\end{align}
Explicitly, the Pauli algebra $\cP$ is given by \eqref{eq:jwsd} with $T_{\!M}$ replaced by $T$, i.e., $\cP\!=\!\fA^{(0)}\!+\!T\fA^{(1)}$, where $\fA^{(1)}$ is the odd subalgebra with respect to the $\Z_{2}$-grading $\Theta$ (see \eqref{eq:grad}). The infinite-length dynamics $\tau_{t}$ of the Hamiltonian $H$ in \eqref{eq:2diH} exists on $\cP$ by general methods for spin systems \cite[Ex.~6.2.14]{BratteliOperatorAlgebrasAnd2} as well as on $\fA$, where it is quasi-free for the self-dual fields, $\tau_{t}(\Psi(\xi,\eta)\!)\!=\!\Psi(e^{ith}(\xi,\eta)\!)$, and given by the diagonal one-particle Hamiltonian,
\begin{align}
\label{eq:2diH1p}
h(\theta) & \!=\! 2\begin{pmatrix} 0 & -i\overline{z_{\theta}} \\ iz_{\theta} & 0 \end{pmatrix}, & \theta & \in[-\pi,\pi),
\end{align}
in momentum space. Thus, to evaluate the dynamical spin-spin correlation functions we essential need to understand the dynamics of the tail operator $T$. Because of the identity,
\begin{align}
\label{eq:eq:2ptjwdyn}
\sigma^{(3)}_{j}\tau_{t}(\sigma^{(3)}_{j'}) & \!=\! (a_{j}\!+\!a^{\dag}_{j})S_{j}T\tau_{t}(T)\tau_{t}(S_{j'}(a_{j'}\!+\!a^{\dag}_{j'})\!),
\end{align}
we essentially need to control the expression $T\tau_{t}(T)$ which is given by perturbation theory for inner automorphisms \cite[Thm.~3.1.33]{BratteliOperatorAlgebrasAnd1}:
\begin{align}
\label{eq:taildyn}
T\tau_{t}(T) & \!=\! \lim_{M\rightarrow\infty}Te^{itH_{\!M}}Te^{-itH_{\!M}} \!=\! \lim_{M\rightarrow\infty}e^{it\Theta_{-}(H_{\!M})}e^{-itH_{\!M}} \!=\! \sum_{n=0}^{\infty}i^{n}\!\!\int_{0}^{t}\!\!dt_{1}\!\!\int_{0}^{t_{1}}\!\!dt_{2}...\!\!\int_{0}^{t_{n-1}}\!\!dt_{n}\!\ \tau_{t_{n}}\!(P)...\tau_{t_{1}}\!(P),
\end{align}
where $P\!=\!\lim_{M\rightarrow\infty}\Theta_{-}(H_{\!M})\!-\!H_{M}\!=\!2t^{(3)}(a_{0}\!-\!a^{\dag}_{0})(a_{1}\!+\!a^{\dag}_{1})$. With a similar strategy, it is possible to control expressions of the form,
\begin{align}
\label{eq:rgtaildyn}
\alpha^{m}(T)\tau_{t}(\alpha^{m}(T)\!) & \!=\! \lim_{M\rightarrow\infty}\lim_{N\rightarrow\infty}e^{it\alpha^{m}(T_{\!N})H_{\!M}\alpha^{m}(T_{\!N})}e^{-itH_{\!M}},
\end{align}
corresponding to perturbations $P^{(m)}\!=\!\lim_{M\rightarrow\infty}\lim_{N\rightarrow\infty}\alpha^{m}(T_{\!N})H_{\!M}\alpha^{m}(T_{\!N})\!-\!H_{\!M}$, appearing in the evaluation of dynamical spin-spin correlation functions.

\section{Error bounds: fermionic correlation functions}
The explicit upper bound on the error $\delta$ in \eqref{eq:error} appearing in the approximation of fermionic correlation functions \eqref{eq:ncor} can be efficiently derived using the self-dual fields $\Psi(\xi,\eta)\!=\!a(\xi\!-\!i\eta)\!+\!a^{\dag}(\overline{\xi\!+\!i\eta})$. It follows from the combinatorial formula for the evaluation of quasi-free states in terms of Pfaffians \eqref{eq:pfaffian} that it is sufficient to derive the bound \eqref{eq:error} for two-point correlations functions which can be evaluated in terms of the one-particle space inner product of $\fh\oplus\fh$ and the one-particle covariance operator $C_{\beta}$ determined by \eqref{eq:KMScov}:
\begin{align}
\label{eq:cov}
C_{\beta} & \!=\! \begin{pmatrix} \1 & i(\1\!-\!2(C^{(1)}_{\beta}\!+\!C^{(2)}_{\beta})\!) \\ -i(\1\!-\!2(C^{(1)}_{\beta}\!+\!C^{(2)}_{\beta})\!) & \1 \end{pmatrix} \!=\! 2(e^{\beta h}+\1)^{-1},
\end{align}
where $h$ is the one-particle Hamiltonian \eqref{eq:2diH1p}. At criticality, i.e, $t^{(1)}\!=\!t^{(3)}\!=\!t$ and $\beta\!\rightarrow\!\infty$, the momentum-space kernel of the covariance reads,
\begin{align}
\label{eq:covcrit}
C_{\!\infty}(\theta,\theta') & \!=\! 2\pi\delta(\theta\!-\!\theta')\underbrace{\begin{pmatrix} 1 & -i\tfrac{\overline{z_{\theta}}}{|z_{\theta}|} \\ i\tfrac{z_{\theta}}{|z_{\theta}|} & 1 \end{pmatrix}}_{=C_{\!\infty}(\theta)} \!=\! 2\pi\delta(\theta\!-\!\theta')\begin{pmatrix} 1 & -i\tfrac{1-e^{-i\theta}}{2|\sin(\frac{1}{2}\theta)|} \\ i\tfrac{1-e^{i\theta}}{2|\sin(\frac{1}{2}\theta)|} & 1 \end{pmatrix}.
\end{align}
which becomes,
\begin{align}
\label{eq:covcritrg}
C(k,k') & \!=\! 2\pi\delta(k\!-\!k')\underbrace{\begin{pmatrix} 1 & \sign(k) \\ \sign(k) & 1 \end{pmatrix}}_{=C(k)},
\end{align}
in the scaling limit at criticality according to \eqref{eq:crit2prg}. Putting everything together, \eqref{eq:ncor} reads as follows for two-point correlation functions of $\Psi$:
\begin{align}
\label{eq:2ptsddyn}
& |\omega^{(m)}\!(\Psi_{\!t^{(\!0\!)}_{1}}\!(\xi_{1},\eta_{1}\!)\Psi_{\!t^{(\!0\!)}_{2}}\!(\xi_{2},\eta_{2}\!)\!)\!-\!\omega(\Psi_{\!t_{1}}\!(\xi_{1},\eta_{1}\!)\Psi_{\!t_{2}}\!(\xi_{2},\eta_{2}\!)\!)| \\ \nonumber
= & |\langle R^{m}\!(\xi_{1},\eta_{1}\!),\!C_{\infty}e^{i(t^{(\!0\!)}_{2}\!-\!t^{(\!0\!)}_{1})h}\!R^{m}\!(\bar{\xi}_{2},\bar{\eta}_{2}\!)\!\rangle\!-\!\langle R^{(\!\infty\!)}\!(\xi_{1},\eta_{1}\!),\!Ce^{i(t_{2}\!-\!t_{1})h^{(\!\infty\!)}}\!R^{(\!\infty\!)}\!(\bar{\xi}_{2},\bar{\eta}_{2}\!)\!\rangle|,
\end{align}
where $R^{(\infty)}\xi\!=\!\xi\ast s$ is the asymptotic one-particle isometry as defined below \eqref{eq:crit2pmaj}, and $h^{(\!\infty\!)}$ is the massless one-particle free-fermion Hamiltonian with momentum-space kernel $h^{(\!\infty\!)}\!(k)\!=\!2k\sigma^{(1)}$. As explained in the main text, we expect the dynamical correlation functions of the $m$-times renormalized lattice model to approximated those of the scaling limit and the associated massless free-fermion dynamics only for effective lattice times of the order of the renormalization scale, i.e., $t^{(0)}_{i}\!=\!2^{m}t_{i}$. Using the Cauchy-Schwarz inequality for Sobolev-type norms (with parameters $\gamma,\gamma_{1},\gamma_{2}\!>\!0$), $\|C_{\infty}\|\!\leq\!2$ (by \eqref{eq:cov}) and the unitarity of the dynamics, we find:
\begingroup
\allowdisplaybreaks
\begin{align}
\label{eq:2ptsddynest}
& |\omega^{(m)}\!(\Psi_{\!t^{(\!0\!)}_{1}}\!(\xi_{1},\!\eta_{1}\!)\Psi_{\!t^{(\!0\!)}_{2}}\!(\xi_{2},\!\eta_{2}\!)\!)\!-\!\omega(\Psi_{\!t_{1}}\!(\xi_{1},\!\eta_{1}\!)\Psi_{\!t_{2}}\!(\xi_{2},\!\eta_{2}\!)\!)| \\ \nonumber
= & |\langle R^{(\!\infty\!)}\!(\xi_{1},\!\eta_{1}\!),\!\Big(\!S_{2^{-m}}C_{\!\infty}e^{i(t^{(\!0\!)}_{2}\!-\!t^{(\!0\!)}_{1})h}\!S_{2^{m}}\!-\!Ce^{i(t_{2}\!-\!t_{1})h^{(\!\infty\!)}}\!\Big)R^{(\!\infty\!)}\!(\bar{\xi}_{2},\!\bar{\eta}_{2}\!)\rangle| \\ \nonumber
= & |\tfrac{1}{2\pi}\int^{\infty}_{-\infty}\!\!\!dk\!\ |\hat{s}(k)|^{2}\begin{pmatrix} \hat{\xi}_{1|k} \\ \hat{\eta}_{1|k} \end{pmatrix}^{\!\dag}\!\!\Big(C_{\!\infty}\!(2^{-m}k)e^{i(t^{(\!0\!)}_{2}\!-\!t^{(\!0\!)}_{1})h(2^{-m}k)}\!-\!C(k)e^{i(t_{2}\!-\!t_{1})h^{\!(\!\infty\!)\!(k)}}\!\Big)\!\!\begin{pmatrix} \hat{\bar{\xi}}_{2|k} \\ \hat{\bar{\eta}}_{2|k} \end{pmatrix}| \\ \nonumber
\leq & |\tfrac{1}{2\pi}\int^{\infty}_{-\infty}\!\!\!dk\!\ |\hat{s}(k)|^{2}\begin{pmatrix} \hat{\xi}_{1|k} \\ \hat{\eta}_{1|k} \end{pmatrix}^{\!\dag}\!\!\Big(C_{\!\infty}\!(2^{-m}k)\!-\!C(k)\!\Big)e^{i(t^{(\!0\!)}_{2}\!-\!t^{(\!0\!)}_{1})h(2^{-m}k)}\!\!\begin{pmatrix} \hat{\bar{\xi}}_{2|k} \\ \hat{\bar{\eta}}_{2|k} \end{pmatrix}| \\ \nonumber
 & \hspace{0.25cm}+\!|\tfrac{1}{2\pi}\int^{\infty}_{-\infty}\!\!\!dk\!\ |\hat{s}(k)|^{2}\begin{pmatrix} \hat{\xi}_{1|k} \\ \hat{\eta}_{1|k} \end{pmatrix}^{\!\dag}\!\!C(k)\Big(e^{i(t^{(\!0\!)}_{2}\!-\!t^{(\!0\!)}_{1})h(2^{-m}k)}\!-\!e^{i(t_{2}\!-\!t_{1})h^{\!(\!\infty\!)\!(k)}}\!\Big)\!\!\begin{pmatrix} \hat{\bar{\xi}}_{2|k} \\ \hat{\bar{\eta}}_{2|k} \end{pmatrix}| \\ \nonumber
 \leq & \tfrac{1}{2\pi}\!\Bigg(\!\int^{\infty}_{-\infty}\!\!\!\!\!dk\!\ \frac{|\hat{s}(k)|^{2(1-\gamma)}}{(1\!+\!|k|^{2})^{\gamma_{2}}}\Big\|\!\Big(\!C_{\!\infty}\!(2^{-m}k)\!-\!C(k)\!\!\Big)\!\!\begin{pmatrix} \hat{\xi}_{1|k} \\ \hat{\eta}_{1|k} \end{pmatrix}\!\!\Big\|^{2}\!\Bigg)^{\!\!\!\frac{1}{2}}\!\!\Bigg(\!\int^{\infty}_{-\infty}\!\!\!\!\!dk\!\ (1\!+\!|k|^{2})^{\gamma_{2}}|\hat{s}(k)|^{2\gamma}\Big\|e^{i(t^{(\!0\!)}_{2}\!-\!t^{(\!0\!)}_{1})h(2^{-m}k)}\!\!\begin{pmatrix} \hat{\xi}_{2|k} \\ \hat{\eta}_{2|k} \end{pmatrix}\!\!\Big\|^{2}\!\Bigg)^{\!\!\!\frac{1}{2}} \\ \nonumber
 & \!\!+\!\tfrac{1}{2\pi}\!\Bigg(\!\int^{\infty}_{-\infty}\!\!\!\!\!dk\!\ (1\!+\!|k|^{2})^{\gamma_{1}}|\hat{s}(k)|^{2(1-\gamma)}\Big\|C(k)\!\!\begin{pmatrix} \hat{\xi}_{1|k} \\ \hat{\eta}_{1|k} \end{pmatrix}\!\!\Big\|^{2}\!\Bigg)^{\!\!\!\frac{1}{2}}\!\!\Bigg(\!\int^{\infty}_{-\infty}\!\!\!\!\!dk\!\ \frac{|\hat{s}(k)|^{2\gamma}}{(1\!+\!|k|^{2})^{\gamma_{1}}}\Big\|\!\Big(\!e^{i(t^{(\!0\!)}_{2}\!-\!t^{(\!0\!)}_{1})h(2^{-m}k)}\!-\!e^{i(t_{2}\!-\!t_{1})h^{\!(\!\infty\!)\!(k)}}\!\Big)\!\!\begin{pmatrix} \hat{\xi}_{2|k} \\ \hat{\eta}_{2|k} \end{pmatrix}\!\!\Big\|^{2}\!\Bigg)^{\!\!\!\frac{1}{2}} \\ \nonumber
 \leq & \tfrac{1}{2\pi}\!\Bigg(\!\int^{\infty}_{-\infty}\!\!\!\!\!dk\!\ \frac{|\hat{s}(k)|^{2(1-\gamma)}}{(1\!+\!|k|^{2})^{\gamma_{2}}}\Big\|\!\Big(\!C_{\!\infty}\!(2^{-m}k)\!-\!C(k)\!\!\Big)\!\!\begin{pmatrix} \hat{\xi}_{1|k} \\ \hat{\eta}_{1|k} \end{pmatrix}\!\!\Big\|^{2}\!\Bigg)^{\!\!\!\frac{1}{2}}\!\!\Bigg(\!\int^{\infty}_{-\infty}\!\!\!\!\!dk\!\ (1\!+\!|k|^{2})^{\gamma_{2}}|\hat{s}(k)|^{2\gamma}\Big\|\!\begin{pmatrix} \hat{\xi}_{2|k} \\ \hat{\eta}_{2|k} \end{pmatrix}\!\!\Big\|^{2}\!\Bigg)^{\!\!\!\frac{1}{2}} \\ \nonumber
 & \!\!+\!\tfrac{2^{\frac{1}{2}}}{2\pi}\!\Bigg(\!\int^{\infty}_{-\infty}\!\!\!\!\!dk\!\ (1\!+\!|k|^{2})^{\gamma_{1}}|\hat{s}(k)|^{2(1-\gamma)}\Big\|\!\begin{pmatrix} \hat{\xi}_{1|k} \\ \hat{\eta}_{1|k} \end{pmatrix}\!\!\Big\|^{2}\!\Bigg)^{\!\!\!\frac{1}{2}}\!\!\Bigg(\!\int^{\infty}_{-\infty}\!\!\!\!\!dk\!\ \frac{|\hat{s}(k)|^{2\gamma}}{(1\!+\!|k|^{2})^{\gamma_{1}}}\Big\|\!\Big(\!e^{i(t^{(\!0\!)}_{2}\!-\!t^{(\!0\!)}_{1})h(2^{-m}k)}\!-\!e^{i(t_{2}\!-\!t_{1})h^{\!(\!\infty\!)\!(k)}}\!\Big)\!\!\begin{pmatrix} \hat{\xi}_{2|k} \\ \hat{\eta}_{2|k} \end{pmatrix}\!\!\Big\|^{2}\!\Bigg)^{\!\!\!\frac{1}{2}} \\ \nonumber
 = & \tfrac{1}{2\pi}\!\Big\|\hat{s}^{2\gamma}\!\!\begin{pmatrix} \hat{\xi}_{2} \\ \hat{\eta}_{2} \end{pmatrix}\!\!\Big\|_{H^{\gamma_{2}}\!(\R)}\!\Bigg(\!\int^{\infty}_{-\infty}\!\!\!\!\!dk\!\ \frac{|\hat{s}(k)|^{2(1-\gamma)}}{(1\!+\!|k|^{2})^{\gamma_{2}}}\Big\|\!\Big(\!C_{\!\infty}\!(2^{-m}k)\!-\!C(k)\!\!\Big)\!\!\begin{pmatrix} \hat{\xi}_{1|k} \\ \hat{\eta}_{1|k} \end{pmatrix}\!\!\Big\|^{2}\!\Bigg)^{\!\!\!\frac{1}{2}} \\ \nonumber
 & \!\!+\!\tfrac{2^{\frac{1}{2}}}{2\pi}\!\Big\|\hat{s}^{2(1-\gamma)}\!\!\begin{pmatrix} \hat{\xi}_{1} \\ \hat{\eta}_{1} \end{pmatrix}\!\!\Big\|_{H^{\gamma_{1}}\!(\R)}\!\Bigg(\!\int^{\infty}_{-\infty}\!\!\!\!\!dk\!\ \frac{|\hat{s}(k)|^{2\gamma}}{(1\!+\!|k|^{2})^{\gamma_{1}}}\Big\|\!\Big(\!e^{i(t^{(\!0\!)}_{2}\!-\!t^{(\!0\!)}_{1})h(2^{-m}k)}\!-\!e^{i(t_{2}\!-\!t_{1})h^{\!(\!\infty\!)\!(k)}}\!\Big)\!\!\begin{pmatrix} \hat{\xi}_{2|k} \\ \hat{\eta}_{2|k} \end{pmatrix}\!\!\Big\|^{2}\!\Bigg)^{\!\!\!\frac{1}{2}}
\end{align}
\endgroup
with $s_{j}(x)\!=\!s(x\!-\!j)$ and $e_{n}$, $n\!=\!1,2$ the standard basis vectors of $\R^{2}$, and where we used the fact the intermediate scaling map, $(R^{(\!\infty\!)}_{m}\hat{f})(k)\!=\!2^{-\frac{m}{2}}\hat{s}(2^{-m}k)\hat{f}(k)$, is an isometry between the Hilbert spaces $L^{2}([-2^{m}\pi,2^{m}\pi),\!(2^{m+1}\pi)^{-1})$ and $L^{2}(\R,\!(2\pi)^{-1})$, and defined the scaling transformation $(S_{\lambda}\hat{f})(k)\!=\!\hat{f}(\lambda k)$ in momentum space. Next, we evaluate the $k$-dependent norms inside the integrals in the last two lines of \eqref{eq:2ptsddynest}:
\begin{align}
\label{eq:knormest}
\Big\|\!\Big(\!C_{\!\infty}\!(2^{-m}k)\!-\!C(k)\!\!\Big)\!\!\begin{pmatrix} \hat{\xi}_{1|k} \\ \hat{\eta}_{1|k} \end{pmatrix}\!\!\Big\|^{2} & \!=\! \Big|\tfrac{i(1-e^{i2^{-m}k})}{2\sin(\frac{1}{2}2^{-m}k)}\!-\!1\Big|^{2}\Big\|\!\!\begin{pmatrix} \hat{\xi}_{1|k} \\ \hat{\eta}_{1|k} \end{pmatrix}\!\!\Big\|^{2}, \\ \nonumber
\Big\|\!\Big(\!e^{i(t^{(\!0\!)}_{2}\!-\!t^{(\!0\!)}_{1})h(2^{-m}k)}\!-\!e^{i(t_{2}\!-\!t_{1})h^{\!(\!\infty\!)\!(k)}}\!\Big)\!\!\begin{pmatrix} \hat{\xi}_{2|k} \\ \hat{\eta}_{2|k} \end{pmatrix}\!\!\Big\|^{2} & \!\leq\! \Big(|\!\cos(2t_{0}t2^{m+1}|\sin(2^{-(m+1)}k)|)\!-\!\cos(2t_{0}t|k|)|  \\ \nonumber
&\hspace{0.5cm}\!+\!|\!\sin(2t_{0}t2^{m+1}|\sin(2^{-(m+1)}k)|)\!-\!\sin(2t_{0}t|k|)| \\ \nonumber
&\hspace{0.5cm}\!+\!|\tfrac{i(1-e^{i2^{-m}k})}{2\sin(\frac{1}{2}2^{-m}k)}\!-\!1|\!+\!|1\!-\!\cos(2^{-m}k)|\Big)^{\!2}\Big\|\!\!\begin{pmatrix} \hat{\xi}_{2|k} \\ \hat{\eta}_{2|k} \end{pmatrix}\!\!\Big\|^{2},
\end{align}
where $t_{0}\!=\!t_{2}\!-\!t_{1}$. Finally, we observe that,
\begin{align}
\label{eq:supest}
\sup_{k\in\R}|k|^{-1}|\tfrac{i(1-e^{ik})}{2\sin(\frac{1}{2}k)}\!-\!1| & \!=\! \tfrac{1}{2}, \\ \nonumber
\sup_{k\in\R}|k|^{-2}|1\!-\!\cos(k)| & \!=\! \tfrac{1}{2}, \\ \nonumber
\sup_{k\in\R}|k|^{-4}|\!\cos(2t_{0}t2^{m}2|\sin(\tfrac{1}{2}k)|)\!-\!\cos(2t_{0}t2^{m}|k|)| & \!=\! 2^{2m}\tfrac{8}{3}(t_{0}t)^{2},  \\ \nonumber
\sup_{k\in\R}|k|^{-3}|\!\sin(2t_{0}t2^{m}2|\sin(\tfrac{1}{2}k)|)\!-\!\sin(2t_{0}t2^{m}|k|)| & \!=\! 2^{m}\tfrac{4}{3}t_{0}t,
\end{align}
which, together with \eqref{eq:2ptsddynest} \& \eqref{eq:knormest}, combine to the final estimate:
\begingroup
\allowdisplaybreaks
\begin{align}
\label{eq:errorest}
 & |\omega^{(m)}\!(\Psi_{\!t^{(\!0\!)}_{1}}\!(\xi_{1},\!\eta_{1}\!)\Psi_{\!t^{(\!0\!)}_{2}}\!(\xi_{2},\!\eta_{2}\!)\!)\!-\!\omega(\Psi_{\!t_{1}}\!(\xi_{1},\!\eta_{1}\!)\Psi_{\!t_{2}}\!(\xi_{2},\!\eta_{2}\!)\!)| \\ \nonumber
  \leq & 2^{-m}\tfrac{2^{\frac{1}{2}}}{2\pi}\!\Bigg(\!\tfrac{2^{\frac{1}{2}}+1}{2^{\frac{1}{2}}}\!\sup_{k\in\R}\!|k|^{-1}|\tfrac{i(1-e^{ik})}{2\sin(\frac{1}{2}k)}\!-\!1|\Big\|\hat{s}^{1\!-\!\gamma}\!\!\begin{pmatrix} \hat{\xi}_{1} \\ \hat{\eta}_{1} \end{pmatrix}\!\!\Big\|_{H^{1}\!(\R)}\Big\|\hat{s}^{\gamma}\!\!\begin{pmatrix} \hat{\xi}_{2} \\ \hat{\eta}_{2} \end{pmatrix}\!\!\Big\|_{H^{1}\!(\R)} \\ \nonumber
 & \hspace{1.5cm}\!+\!2^{-m}\sup_{k\in\R}|k|^{-2}|1\!-\!\cos(k)|\Big\|\hat{s}^{1\!-\!\gamma}\!\!\begin{pmatrix} \hat{\xi}_{1} \\ \hat{\eta}_{1} \end{pmatrix}\!\!\Big\|_{H^{2}\!(\R)}\Big\|\hat{s}^{\gamma}\!\!\begin{pmatrix} \hat{\xi}_{2} \\ \hat{\eta}_{2} \end{pmatrix}\!\!\Big\|_{H^{2}\!(\R)} \\ \nonumber
 & \hspace{1.5cm}\!+\!2^{-2m}\sup_{k\in\R}|k|^{-3}|\!\sin(2t_{0}t2^{m}2|\sin(\tfrac{1}{2}k)|)\!-\!\sin(2t_{0}t2^{m}|k|)|\Big\|\hat{s}^{1\!-\!\gamma}\!\!\begin{pmatrix} \hat{\xi}_{1} \\ \hat{\eta}_{1} \end{pmatrix}\!\!\Big\|_{H^{3}\!(\R)}\Big\|\hat{s}^{\gamma}\!\!\begin{pmatrix} \hat{\xi}_{2} \\ \hat{\eta}_{2} \end{pmatrix}\!\!\Big\|_{H^{3}\!(\R)} \\ \nonumber
 & \hspace{1.5cm}\!+\!2^{-3m}\sup_{k\in\R}|k|^{-4}|\!\cos(2t_{0}t2^{m}2|\sin(\tfrac{1}{2}k)|)\!-\!\cos(2t_{0}t2^{m}|k|)|\Big\|\hat{s}^{1\!-\!\gamma}\!\!\begin{pmatrix} \hat{\xi}_{1} \\ \hat{\eta}_{1} \end{pmatrix}\!\!\Big\|_{H^{4}\!(\R)}\Big\|\hat{s}^{\gamma}\!\!\begin{pmatrix} \hat{\xi}_{2} \\ \hat{\eta}_{2} \end{pmatrix}\!\!\Big\|_{H^{4}\!(\R)}\!\Bigg) \\ \nonumber
 \leq & 2^{-m}\tfrac{2^{\frac{1}{2}}}{2\pi}\!\Bigg(\!\tfrac{2^{\frac{1}{2}}+1}{2^{\frac{1}{2}}2}\Big\|\hat{s}^{1\!-\!\gamma}\!\!\begin{pmatrix} \hat{\xi}_{1} \\ \hat{\eta}_{1} \end{pmatrix}\!\!\Big\|_{H^{1}\!(\R)}\Big\|\hat{s}^{\gamma}\!\!\begin{pmatrix} \hat{\xi}_{2} \\ \hat{\eta}_{2} \end{pmatrix}\!\!\Big\|_{H^{1}\!(\R)}\!+\!2^{-m}\tfrac{1}{2}\Big\|\hat{s}^{1\!-\!\gamma}\!\!\begin{pmatrix} \hat{\xi}_{1} \\ \hat{\eta}_{1} \end{pmatrix}\!\!\Big\|_{H^{2}\!(\R)}\Big\|\hat{s}^{\gamma}\!\!\begin{pmatrix} \hat{\xi}_{2} \\ \hat{\eta}_{2} \end{pmatrix}\!\!\Big\|_{H^{2}\!(\R)} \\ \nonumber
 & \hspace{1.5cm}\!+\!2^{-m}\tfrac{4}{3}t_{0}t\Big\|\hat{s}^{1\!-\!\gamma}\!\!\begin{pmatrix} \hat{\xi}_{1} \\ \hat{\eta}_{1} \end{pmatrix}\!\!\Big\|_{H^{3}\!(\R)}\Big\|\hat{s}^{\gamma}\!\!\begin{pmatrix} \hat{\xi}_{2} \\ \hat{\eta}_{2} \end{pmatrix}\!\!\Big\|_{H^{3}\!(\R)}\!+\!2^{-m}\tfrac{8}{3}(t_{0}t)^{2}\Big\|\hat{s}^{1\!-\!\gamma}\!\!\begin{pmatrix} \hat{\xi}_{1} \\ \hat{\eta}_{1} \end{pmatrix}\!\!\Big\|_{H^{4}\!(\R)}\Big\|\hat{s}^{\gamma}\!\!\begin{pmatrix} \hat{\xi}_{2} \\ \hat{\eta}_{2} \end{pmatrix}\!\!\Big\|_{H^{4}\!(\R)}\!\Bigg) \\ \nonumber
\leq & 2^{-m}C_{T}\!\!\!\!\!\!\!\!\!\max_{\gamma_{1}\!+\!\gamma_{2}\in\{1,2,3,4\}}\!\Big\|\hat{s}^{1\!-\!\gamma}\!\!\begin{pmatrix} \hat{\xi}_{1} \\ \hat{\eta}_{1} \end{pmatrix}\!\!\Big\|_{H^{\gamma_{1}}\!(\R)}\Big\|\hat{s}^{\gamma}\!\!\begin{pmatrix} \hat{\xi}_{2} \\ \hat{\eta}_{2} \end{pmatrix}\!\!\Big\|_{H^{\gamma_{2}}\!(\R)},
\end{align}
\endgroup
for $|t_{0}|\!\in\![0,T]$, a constant $C_{T}\!>\!0$ and some free parameter $\gamma\!>\!0$. We note that the Sobolev-type norm are finite for sufficiently regular scaling functions $s\!\in\!C^{r}\!(\R)$. Moreover, inspecting the penultimate line of \eqref{eq:errorest} closely, we find that we can even choose the maximal continuum time to scale as $T\!\lesssim\!2^{m}T_{0}$ for some $T_{0}\!>\!0$, i.e., the approximation error $\delta\!=\!\delta(m,T)$ vanishes in the limit $m\!\rightarrow\!\infty$ as long as $2^{-m}T\!\rightarrow\!0$. In particular, the effective lattice times $t^{(\!0\!)}_{i}$ only need to be asymptotically large compared to the continuum times $t_{i}$, at the cost of an approximation error $\delta$ vanishing at least with the inverse asymptotics.

\end{document}